\documentclass[pre,aps,showpacs,twocolumn]{revtex4}
\usepackage{amsfonts}
\usepackage{amsmath}
\usepackage{amssymb}
\usepackage{graphicx}
\usepackage{epsfig}
\usepackage{bm}
\usepackage{multirow}

\begin{document}

\title{Generic Two-Qubit Photonic Gates Implemented by Number-Resolving Photodetection}

\author{Dmitry B. Uskov, A. Matthew Smith, and Lev Kaplan}%
\affiliation{Department of Physics, Tulane University, New Orleans, Louisiana 70118, USA}%


\begin{abstract}
We combine numerical optimization techniques [Uskov et al., Phys. Rev. A {\bf 79}, 042326 (2009)] with symmetries
of the Weyl chamber
to obtain optimal implementations of generic linear-optical KLM-type two-qubit entangling gates.
We find that while any two-qubit controlled-U gate, including {\sc{cnot}} and {\sc{cs}}, can be implemented using only two ancilla resources with success probability $S>0.05$, a generic $SU(4)$ operation requires three unentangled ancilla photons,
with success $S>0.0063$.
Specifically, we obtain a maximal success probability close to 0.0072 for the B gate.
We show that single-shot implementation of a generic $SU(4)$ gate
offers more than an order of magnitude increase in the success probability and two-fold reduction in overhead ancilla resources compared to standard triple-{\sc{cnot}} and double-B gate decompositions. 
\end{abstract}


\maketitle
Generation of and operation on quantum states of light at the single-photon level are important topics of research in the field of theoretical and experimental quantum information and metrology \cite{kwiat,kimble,lanyon}. Due to the low rate of decoherence, photonic states are capable of carrying quantum information over large distances, enabling quantum state teleportation and distribution of entanglement across quantum networks.
To build a universal all-optical quantum computer, one may couple photons through their interaction with atomic media resulting either in nonlinear but unitary interaction~\cite{aoki} or in two-photon dissipative coupling and Zeno-type non-unitary evolution~\cite{franson}. However, nonlinear effects are vanishingly small for field intensities at the single-photon level, and the feasibility of these approaches is still unclear.

\begin{figure}[htbp]
   \includegraphics[width=0.42\textwidth]{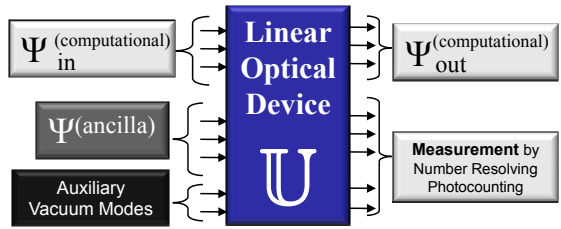} 
   \caption{(Color online) A scheme for an LOQC transformation. The computational input state is a separable state of two or more dual-rail encoded qubits. The ancilla state may be a separable state, an entangled state, or an ebit state carrying spatially distributed entanglement~\cite{wilde}.}
\label{figschematic}
\end{figure}

A more explicit and straightforward solution to the problem of photon coupling was suggested in a seminal work by Knill, Laflamme, and Milburn~\cite{knill}. In the KLM scheme, linear optical operations are performed on photons in computational and ancilla modes, followed by a measurement of the ancilla modes using number-resolving photocounting as shown in Fig.~\ref{figschematic}.

Importantly,
the bosonic statistics of photons, and the availability of experimental sources for generating photons indistinguishable in the spatial degrees of freedom, allow the quantum Hong-Ou-Mandel effect~\cite{hong} to be exploited. The fidelity of such transformations strongly depends on mode mismatch
(for experimental details of the errors involved, see for example Ref.~\cite{lanyon}).
However, recent progress in the technology of manufacturing microchips for optical interferometers, and the improvement of down-conversion sources of entangled photon pairs~\cite{kwiat} inspires strong optimism that linear optical quantum computation (LOQC) will become a practical quantum technology in the near future.

An LOQC measurement-assisted transformation is schematically illustrated in Fig. 1.
The total input state is $|\Psi_{\rm in}^{\rm (total)}\rangle=|{\Psi}_{\rm in}^{\rm (comp)}\rangle \otimes | \Psi^{\rm (ancilla)}\rangle$, where $|{\Psi}_{\rm in}^{\rm (comp)}\rangle$ is a computational state in $N_c$ modes and $| \Psi^{\rm (ancilla)}\rangle$ is an ancilla state in $N_a$ modes. In the Fock representation, this input state is determined by occupation numbers $n_1, \ldots, n_{N_c}$ and $n_{N_c+1},\ldots,n_{N_c+N_a}$ for the computational modes and ancilla modes, respectively.

As indicated in Fig.~\ref{figschematic}, the linear optical device acts on the modes by a unitary transformation 
$a_i^\dag \to \sum_{j=1}^N U_{ij} \tilde{a}^\dag_j$~\cite{perelomov}, where $U$ is a unitary $N\times N$ matrix associated with the concrete optical device, $a^\dag$ and $\tilde{a}^\dag$ are creation operators of the input and output modes, respectively, and $N$ is the total number of modes (including vacuum modes, if any).
Then $|\Psi_{\rm in}^{\rm (total)}\rangle$ is transformed as
\begin{equation}
|\Psi_{\rm out}^{\rm (total)}\rangle={\bm \Omega}(U)|\Psi_{\rm in}^{\rm (total)}\rangle=
\prod_{i=1}^N {1\over{\sqrt{{\it n_i}!}}}\left(\sum_j {\it U_{ij}} \tilde{a}^\dag_j\right )^{n_i}|0\rangle \,,
\label{omega}
\end{equation}
where $\bm \Omega$ is a homogeneous polynomial function in entries of the matrix $U$~\cite{perelomov,vanmeter}.

Ideally, the photocounting measurement projects $|\Psi_{\rm out}^{\rm (total)}\rangle$ onto some state $\langle  k_{N_c+1},k_{N_c+2},\ldots,k_{N_c+N_a}|$ in ancilla modes~\cite{helstrom}. The computational state (unnormalized) becomes
\begin{equation}
\begin{array}{cc}
|\Psi_{\rm out}^{\rm (comp)}\rangle=\langle  k_{N_c+1},k_{N_c+2},...,k_{N}|{\bm \Omega}(U)|\Psi_{\rm in}^{\rm (total)}\rangle \\
 ={\bm  A}(U)|{\Psi}_{\rm in}^{\rm (comp)}\rangle \;\;\;\;\;\;\;\;\;\;\;\;\;\;\;\;\;\;
\end{array}
\label{atransf}
\end{equation}
where ${\bm A}={\bm A}(U)$ is a Kraus operator~\cite{kraus}, $\|{\bm A}\| \leq 1$. All relevant properties of the LOQC transformation (\ref{atransf})
are determined by the matrix ${\bm A}$, whose entries are again homogeneous polynomials in the elements of $U$
(specifically, they are permanents of $U$~\cite{vanmeter}).

Denoting the desired transformation matrix by ${\bm A}^{(\rm tar)}$, we can define the fidelity $F$ and operational success probability $S$ of the transformation (\ref{atransf}) via the Hilbert-Schmidt scalar product $\langle A| B\rangle \equiv {\rm Tr}( AB^\dag)/D_c$ in the ${D_c}$-dimensional computational Hilbert space as follows~\cite{uskov}:
\begin{equation}
\begin{array}{cc}
{F(U)}=|\langle{\bm A}|{\bm A^{(\rm tar)}}\rangle|^2
 / {\langle{\bm A}|{\bm A}\rangle \langle{\bm A^{(\rm tar)}}|{\bm A^{(\rm tar)}}\rangle}\\
\\
{S(U)= \langle{\bm A}|{\bm A}\rangle}\,.
\label{fs}
\end{array}
\end{equation}
Since the success of the LOQC gate operation hinges on the outcome of a stochastic measurement process, the optimization of success probability $S$ using minimal resources while maintaining perfect fidelity $F=1$
is the key problem for practical implementation of LOQC.
This optimization problem does not have a general algebraic solution due to the algebraic complexity of the fidelity and success functions. Even in the simplest case of a {\sc{cnot}} gate, the analytic solution was identified using numerical methods, but never proved analytically to be a global maximum~\cite{knill2,uskov}. The only known example of an analytically optimizable LOQC gate is the Nonlinear Sign ({\sc{ns}}) gate, where convexity of the success probability function allows for global optimization in the restricted case of an unentangled ancilla resource~\cite{eisert}. Even the problem of numerically calculating the coefficients of $\bm A$ belongs to the \#P-complete complexity class~\cite{valiant}.

Recently, several successful implementations of analytical and numerical optimization have been reported in the literature for basic gates, including the {\sc{cnot}} gate~\cite{uskov}, Toffoli gate~\cite{lanyon,ralph,uskov}, and Fredkin gate~\cite{gong}. One important result of these works was a demonstration that LOQC transformations allow one to bypass the standard circuit paradigm of quantum computing, in which the complete quantum calculation is constructed as a product of concatenated standard two-qubit gates. For example, the standard circuit scheme for implementing the Toffoli gate using six {\sc{cnot}} gates~\cite{nielsen} results in a very small success probability of $(2/27)^6$, whereas only three two-qubit gates are required in the analytic scheme~\cite{ralph}, with a success probability of $(2/27)^2/2$ (here we assume a non-entangled ancilla resource and non-destructive heralded implementation of the gate). Further improvement in the success probability is obtained by a single-shot numerical optimization technique~\cite{uskov}, where the target gate is ${\it not}$ decomposed into two-qubit gates but, instead, implemented as a single LOQC transformation.

Universal two-qubit unitary gates constitute the core of current schemes for quantum information processing,
since an arbitrary
$SU(2^n)$ unitary operation can be implemented as a series of two-particle transformations.
The success of the single-shot (block-optimization) technique applied to {\sc{cnot}} and Toffoli transformations inspired us to further investigate the problem of minimizing overhead resources and maximizing success probabilities for an arbitrary $SU(4)$ two-qubit gate.

Until now, the only two-qubit photonic gate systematically studied in the literature was the {\sc{cs}} gate, equivalent to the {\sc{cnot}} gate via local Hadamard transformations, and consequently represented by the same point $\{\pi/2,0,0\}$ in the Weyl chamber (see below). Both gates belong to the class of two-qubit controlled-unitary gates $C^1U$. The {\sc{cnot}} gate is indeed one of the most universal gates, since an arbitrary $SU(4)$ gate may be
constructed using three {\sc{cnot}} gates~\cite{nielsen}. However, {\sc{cnot}} is less universal than the B gate, discovered recently by Zhang et al.~\cite{zhang}. An arbitrary $SU(4)$ transformation can be constructed as a product of only two B gates (plus local qubit rotations).

The success probability of the {\sc{cnot}} gate and required photonic resources are well established~\cite{uskov,knill2}. {\sc{cnot}} requires two unentangled ancilla photons, and the maximal success probability is $2/27\approx0.074$. Surprisingly, adding ancilla resources does not affect the success probability~\cite{uskov}. Consequently, six unentangled ancilla photons will be required to implement a generic two-qubit transformation and the success probability of such a transformation will be rather small, $S=(2/27)^3$. In this work we optimize generic two-qubit gates directly~\cite{uskov}, including the B gate as a special case. This will allow us to determine whether a combination of two B gates (or three {\sc{cnot}} gates) is more efficient than direct implementation of a generic two-qubit coupling gate.

Our method is based on conjugate gradient algorithms for maximizing the fidelity and success probability functions (\ref{fs}) in the space of unitary matrices $U$. Fidelity is a differentiable (but nonanalytic) function of the entries of $U$, defined by equation (\ref{fs}) even when the domain of matrices $U$ is extended by relaxing the unitarity requirement in favor of improving the efficiency of numerical optimization (unitarity is then easily recovered by a unitary dilation procedure if the resulting optimal matrix is found to be non-unitary). The success probability function $S$ is differentiable in the space of unitary matrices $U$, but exhibits singular behavior as a function of $U$ in the extended search space. The optimization starts by optimizing the fidelity $F$ until a point of perfect fidelity $F=1$ is identified, and a penalty function approach is then used to optimize $S$ in the vicinity of the $F=1$ subspace, finally leading to a local maximum of $S$ within the $F=1$ subspace (values of $F$ obtained numerically are better than $0.999999$). The process is repeated with multiple random starting points $U$ to obtain the best success rate for a given target gate.

\begin{figure}[htbp]
   \includegraphics[width=0.33\textwidth]{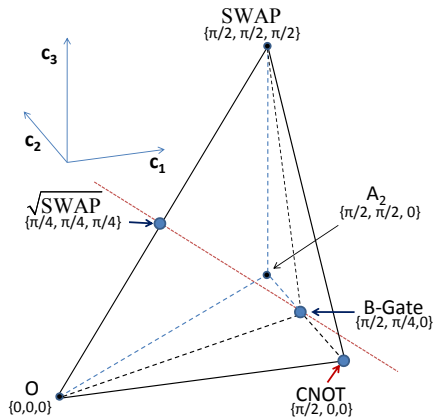} 
   \caption{(Color online) Half of the Weyl Chamber for the Cartan decomposition of the $SU(4)$ group.}
   \label{figchamber}
\end{figure}

A generic two-qubit transformation $V\in SU(4)$ can be implemented as a product of local single-qubit pre- and post-rotations $X_{\rm pre}^{(1)}, X_{\rm pre}^{(2)}, X_{\rm post}^{(1)}, X_{\rm post}^{(2)}\in SU(2)$, and an entangling operation characterized by only three real parameters $\{c_1,c_2,c_3\}$~\cite{khaneja,zhang2}. This is
known as the Cartan KAK decomposition:
\begin{equation}
V=X_{\rm post}^{(1)}X_{\rm post}^{(2)} e^{\frac{i}{2} (c_1\sigma_x \sigma_x+c_2\sigma_y \sigma_y+c_3\sigma_z \sigma_z)}X_{\rm pre}^{(1)}X_{\rm pre}^{(2)} \,.
\end{equation}
Two gates are equivalent up to local rotations plus $\pi$-shifts if and only if the triplets $\{c_1^a,c_2^a,c_3^a\}$ and $\{c_1^b,c_2^b,c_3^b\}$ can be transformed into each other by the action of the Weyl group. These transformations
are listed explicitly in the first three rows of Table~\ref{transformtable}, along with the local rotations generating them. The resulting equivalence class in the space of $\{c_1,c_2,c_3\}$ is known as the Weyl chamber~\cite{zhang2}: $0 \le c_3 \le c_2 \le c_1 \le \pi-c_2$. Half of the Weyl chamber ($c_1<\pi/2$) is shown in Fig.~\ref{figchamber}, with several important gates identified explicitly.

\begin{table}[htbp]
\begin{center}
\begin{tabular}{| c |c |}
\hline
Weyl Symmetry  & Local Rotation \\ 
\hline
$\{c_1,c_2,c_3\}\leftrightarrow\{c_1,\pm c_3,\pm c_2\}$ & $\exp{(\frac{i\pi}{4}(\sigma_x^1\pm \sigma_x^2))}$\\
\hline
 $\{c_1,c_2,c_3\}\leftrightarrow \{\pm c_2,\pm c_1,c_3\}$ & $\exp{(\frac{i\pi}{4}(\sigma_y^1\pm \sigma_y^2))}$\\
\hline
$\{c_1,c_2,c_3\}\leftrightarrow\{\pm c_3,c_2,\pm c_1\}$ &  $\exp{(\frac{i\pi}{4}(\sigma_z^1\pm \sigma_z^2))}$  \\
\hline \hline
Other Symmetry  &  Transformation \\ \hline
$\{c_1,c_2,c_3\}\leftrightarrow\{\pi-c_1,c_2,c_3\}$ & Conjugation + Local \\
\hline
$\{c_1,c_2,c_3\}\leftrightarrow\{\frac{\pi}{2}\!-\!c_3,\frac{\pi}{2}\!-\!c_2,\frac{\pi}{2}\!-\!c_1\}$
& {\sc{swap}} + Conj + Local \\
\hline
\end{tabular}
\end{center}
\caption{Transformations on $\{c_1,c_2,c_3\}$ that preserve the success probability $S$. The first three
transformations are generated by local qubit rotations, as shown, and define the Weyl chamber. Two additional symmetries are specific to LOQC and allow the space of $\{c_1,c_2,c_3\}$ to be reduced to one quarter of the Weyl chamber.}
\label{transformtable}
\end{table}


\begin{figure}[htbp]
   \includegraphics[width=0.23\textwidth]{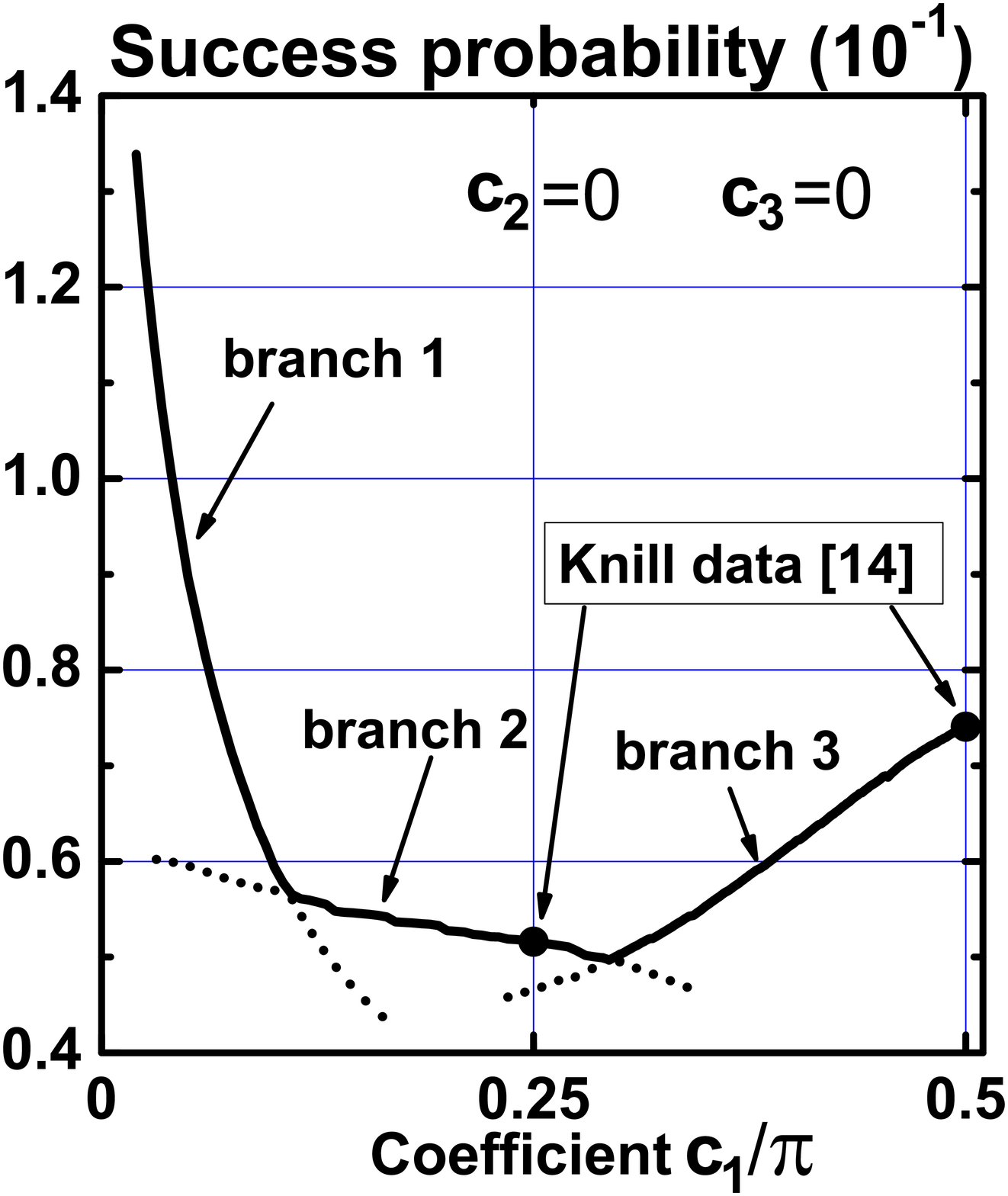}
   \includegraphics[width=0.23\textwidth]{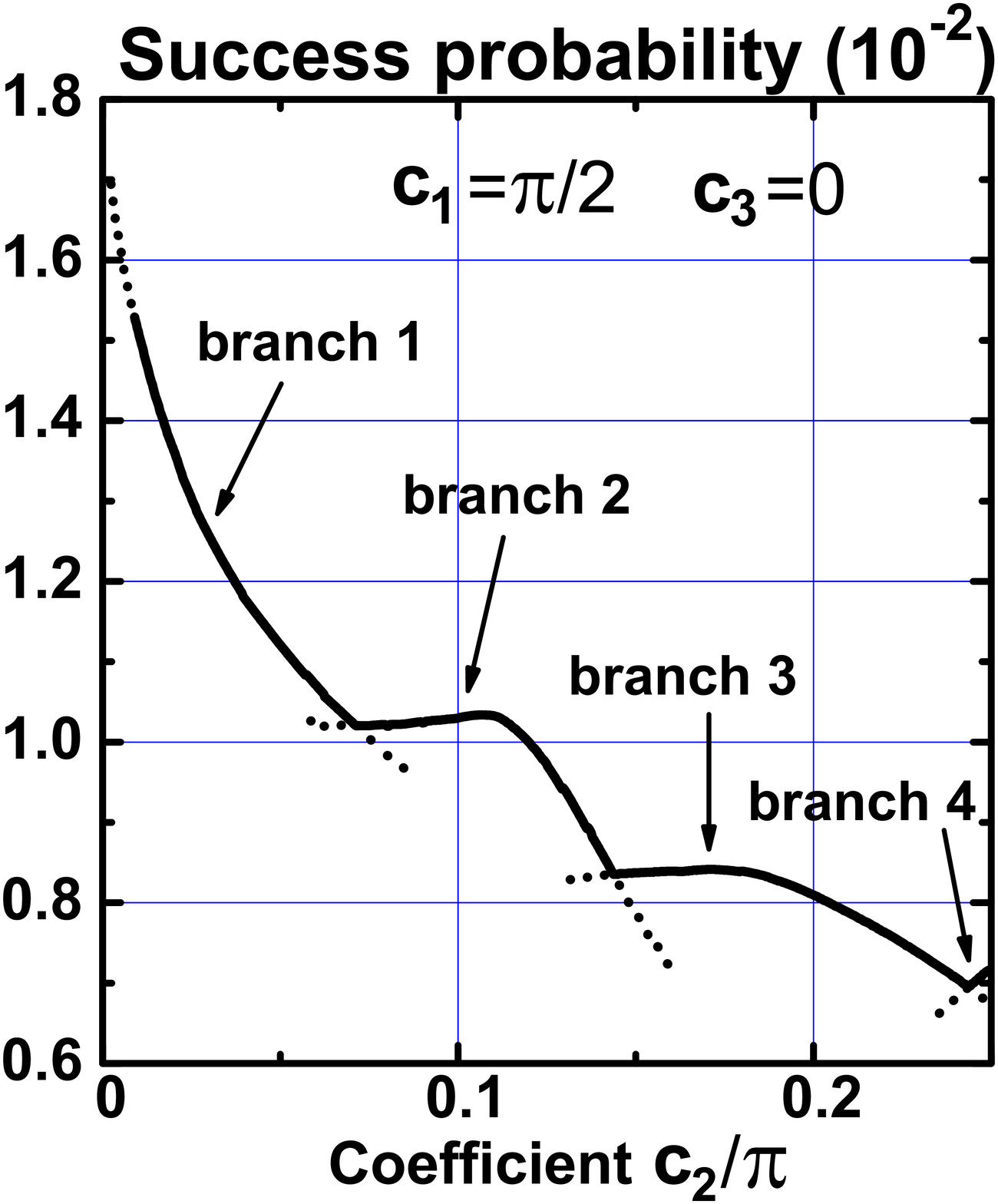}
   \caption{Left panel: Success probability for the $C^1U$ gates along the O-{\sc{cnot}} line ($c_2=c_3=0$), as a function of $c_1$. Right panel: Success probability for gates along the {\sc{cnot}}-B line ($c_1=\pi/2$, $c_3=0$), as a function of $c_2$. Dotted curves indicate continuations of optimal families of solutions.}
   \label{figsuccess}
\end{figure}

Two additional symmetries are present in the problem, allowing us to restrict our attention to one quarter of the Weyl
chamber. First, suppose that a unitary transformation $U$ acting on the photon modes produces a Kraus operator
${\bm A}(U)$ acting on the computational input state, and implements a two-qubit gate $V=e^{\frac{i}{2} (c_1\sigma_x \sigma_x+c_2\sigma_y \sigma_y+c_3\sigma_z \sigma_z)}$ with fidelity $F(U)$ and success probability $S(U)$. From
Eqs.~(\ref{omega}), (\ref{atransf}), it is evident that ${\bm A}(U^\ast)={\bm A}(U)^\ast$, and thus $U^\ast$ implements the gate $V^\ast$ with the same fidelity and success rate. Now the gate $V^\ast$ is associated with the triplet $\{-c_1,-c_2,-c_3\}$, which via local transformations and $\pi$-shifts maps to $\{\pi-c_1,c_2,c_3\}$ in the Weyl
chamber. Thus, $\{c_1,c_2,c_3\}$ and $\{\pi-c_1,c_2,c_3\}$ have the same maximal success probability.
This symmetry corresponds geometrically to a reflection through the {\sc{cnot}}-${\rm A}_2$-{\sc{swap}} plane, and permits us to 
consider only the left half of the Weyl chamber: $0 \le c_3 \le c_2 \le c_1 \le \pi/2$, which is shown in Fig.~\ref{figchamber}.

Secondly, we notice that the {\sc{swap}} operation $\{\pi/2,\pi/2,\pi/2\}$ corresponds simply to permutation of the photon modes, and may always be implemented with perfect fidelity and success. Thus the maximal success for $\{c_1,c_2,c_3\}$ is the same as the maximal success for $\{c_1+\pi/2,c_2+\pi/2,c_3+\pi/2\}$, which by local rotations and a $\pi$-shift maps to $\{\pi/2-c_3,\pi/2-c_2,\pi/2-c_1\}$ inside the half-chamber obtained previously. This last symmetry corresponds to reflection through the B-$\sqrt{\textsc{swap}}$ line in Fig.~\ref{figchamber}, and allows us
to focus on a quarter chamber defined, for example, by vertices O, {\sc{cnot}}, ${\rm A}_2$, and $\sqrt{\textsc{swap}}$.

Each point inside this region of the Weyl chamber represents a distinct gate with its own maximal success probability and minimal number of ancillary photons required to implement it. Gates along the O-{\sc{cnot}} edge are equivalent to the controlled unitary gates $C^1U$~\cite{zhang3}. We find that all these gates require only two ancilla resources to attain prefect fidelity. The maximal success probability as a function of $c_1$ is shown in Fig.~\ref{figsuccess} (left panel). Interestingly, the optimal solution in each case takes the form
previously observed by Knill for $c_1=\pi/2$ ({\sc{cnot}}) and $c_1=\pi/4$~\cite{knill2}, i.e., the $6 \times 6$ $U$ matrix acts trivially on two of the computational modes. 

The optimization process may be aided substantially by considering an $F=1$ solution obtained for
a given gate $\{c_1,c_2,c_3\}$ and using it at a starting point for optimization of nearby gates $\{c'_1,c'_2,c'_3\}$.
This procedure results in a family of locally optimal solutions in a region of the Weyl chamber. Of course, the
continuation of a globally optimal solution at $\{c_1,c_2,c_3\}$ is not guaranteed to remain globally optimal forever,
and one must in general consider multiple such families. As seen in Fig.~\ref{figsuccess} (left panel), the optimal success curve for the O-{\sc{cnot}} line consists of only three distinct families of solutions.

\begin{table}[htbp]

\begin{center}
\begin{tabular}{| c |c |c |}
\hline
\multirow{3}{*} & Success & Ancilla \\
& $S$ & Photons \\
\hline
\multirow{1}{*} 
3 {\sc{cnot}} gates & $(2/27)^3 \approx 0.000406$ & 6 \\
\hline
2 B gates & $>(0.00717)^2 \approx 0.00005$ & 6 \\
\hline
Single shot & $>0.0063$ &  3 \\
\hline
\end{tabular}
\end{center}
\caption{Success probabilities and needed resources for implementing general two-qubit gates, using {\sc{cnot}}
decomposition, B decomposition, and single-shot design.}
\label{successtable}
\end{table}

Away from the O-{\sc{cnot}} line (or the ${\rm A}_2$-{\sc{swap}} line, which is equivalent to it), all two-qubit gates require three ancillas to obtain $F=1$ with $S>0$. A systematic investigation of all non-equivalent gates in the Weyl chamber, on a cubic lattice with spacing $\pi/16$ in $\{c_1,c_2,c_3\}$ space, shows that all $SU(4)$ gates on this
lattice
may be implemented with perfect fidelity and a success probability not lower than $0.0063$. Again, the solutions may be
classified into a number of families. In the right panel of Fig.~\ref{figsuccess}, we show the best success probability
obtained for gates lying along the {\sc{cnot}}-B line (not including the {\sc{cnot}} gate itself,
which requires fewer resources). As indicated in the figure, four families of solutions are found to be globally optimal at different points along this edge of the Weyl chamber. The B gate has success probability $\approx 0.0072$.


Our main result is summarized in Table~\ref{successtable}, which shows the success probability and resources required to implement a generic two-qubit $SU(4)$ gate using three {\sc{cnot}} gates, two B gates, or single-shot design.

In conclusion, we use numerical optimization techniques~\cite{uskov} to find optimal implementations of generic linear-optical KLM-type two-qubit entangling gates, represented by generic points in the Weyl chamber of Khaneja's KAK decomposition of the $SU(4)$ group. Symmetries of the Weyl chamber are identified, and used to aid the
optimization process. A solution at one point in the Weyl chamber may be continuously deformed to obtain a family of locally optimal solutions; several such families are needed to obtain globally optimal solutions at all points in the Weyl chamber. We find that while any two-qubit controlled-U gate, including {\sc{cnot}}  and {\sc{cs}}, can be implemented using only two ancilla resources with success probability $S>0.05$, a generic $SU(4)$ operation requires three unentangled ancilla photons. Our study indicates that single-shot implementation of a generic $SU(4)$ gate
offers more than an order of magnitude increase in the success probability and two-fold reduction in overhead ancilla resources compared to standard triple-{\sc{cnot}}  and double-B gate decompositions. 
The B gate, which is the most efficient deterministic gate for decomposing an arbitrary $SU(4)$ transformation, has success probability close to 0.0072. In the context of probabilistic KLM-type transformations, this makes the B gate less efficient than the {\sc{cnot}}  gate as a building block for arbitrary $SU(4)$ transformations. Our results are consistent with previous work on the Deutsch-Toffoli gate, where direct implementation of this three-qubit operation was shown to be four orders of magnitude more efficient than six-fold decomposition into {\sc{cnot}}  gates \cite{ralph,uskov}.

We thank A. Gilchrist, J. Vala, and M. M. Wilde for very helpful discussions.
This work was supported in part by the NSF under Grants PHY05-51164 and PHY-0545390.


\begin{thebibliography}{99}

\bibitem{kwiat} P.~G.~Kwiat,
Nature {\bf 453}, 294 (2008).

\bibitem{kimble} H.~J.~Kimble,
Nature {\bf 453}, 1023 (2008).

\bibitem{lanyon} B.~P.~Lanyon, M.~A.~Barbieri, M.~P.~Almeida, T.~Jennewein, T.~C.~Ralph, K.~J.~Resch, G.~J.~Pryde, J.~L.~O'Brien, A.~Gilchrist, and A.~G.~White,
Nature Physics {\bf 5}, 134 (2009).

\bibitem{aoki} T.~Aoki, A.~S.~Parkins, D.~J.~Alton, C.~A.~Regal, B.~Dayan, E.~Ostby, K.~J.~Vahala, and H.~J.~Kimble,
Phys. Rev. Lett. {\bf 102}, 083601 (2009).
 
\bibitem{franson} J.~D.~Franson, B.~C.~Jacobs, and T.~B.~Pittman,
Phys. Rev. A {\bf 70}, 062302 (2004).

\bibitem{knill} E.~Knill, R.~Laflamme, and G.~J.~Milburn,
Nature {\bf 409}, 46 (2001).


\bibitem{wilde} M.~M.~Wilde and D.~B.~Uskov,
Phys. Rev. A {\bf 79}, 022305 (2009).

\bibitem{hong} C.~K.~Hong, Z.~Y.~Ou, and L.~Mandel,
Phys. Rev. Lett. {\bf 59}, 2044 (1987).

\bibitem{perelomov} A.~Perelomov,
{\it Generalized Coherent States and Their Applications} (Springer, Berlin, 1986).


\bibitem{vanmeter} N.~M.~VanMeter, P.~Lougovski, D.~B.~Uskov, K.~Kieling, J.~Eisert, and J.~P.~Dowling,
Phys. Rev. A {\bf 76}, 063808 (2007).

\bibitem{helstrom} C.~W.~Helstrom, {\it Quantum detection and estimation theory} (Academic Press, New York, 1976).

\bibitem{kraus} K.~Kraus, Lecture Notes: {\it States, Effects and Operations: Fundamental Notions of Quantum Theory} (Springer, New York, 1983).

\bibitem{uskov} D.~B.~Uskov, L.~Kaplan, A.~M.~Smith, S.~D.~Huver, and J.~P.~Dowling,
Phys. Rev. A {\bf 79}, 042326 (2009).

\bibitem{knill2} E. Knill,
Phys. Rev. A {\bf 66}, 052306 (2002).
 
\bibitem{eisert} J.~Eisert,
Phys. Rev. Lett. {\bf 95}, 040502 (2005).
 
\bibitem{valiant} L.~G.~Valiant,
Theor. Comp. Sci. (Elsevier) {\bf 8}, 189 (1979).
 
\bibitem{ralph} T.~C.~Ralph, K.~J.~Resch, and A.~Gilchrist,
Phys. Rev. A {\bf 75}, 022313 (2007).

\bibitem{gong} Y.-X.~Gong, G.-C.~Guo, and T.~C.~Ralph,
Phys. Rev. A {\bf 78}, 012305 (2008).

\bibitem{nielsen} M.~A.~Nielsen and I.~L.~Chuang,
{\it Quantum Computation and Quantum Information} (Cambridge University Press, Cambridge, 2000); T. Sleator and H. Weinfurter,
Phys. Rev. Lett. {\bf 74}, 4087 (1995).

\bibitem{zhang} J.~Zhang, J.~Vala, S.~Sastry, and K.~B.~Whaley,
Phys. Rev. Lett. {\bf 93}, 020502 (2004).

\bibitem{khaneja} N.~Khaneja and S.~J.~Glaser,
Chem. Phys. {\bf 267}, 11 (2001).

\bibitem{zhang2} J.~Zhang, J.~Vala, S.~Sastry, and K.~B.~Whaley,
Phys. Rev. A {\bf 67}, 042313 (2003).

\bibitem{zhang3} J.~Zhang, J.~Vala, S.~Sastry, and K.~B.~Whaley,
Phys. Rev. A {\bf 69}, 042309 (2004).


\end{thebibliography}
\end{document}